\documentclass{elsart}

\usepackage{graphicx}
\graphicspath{{fig/}}

\usepackage{amssymb}

\begin{document}

\begin{frontmatter}



\title{Performance of a Large Area Avalanche Photodiode in a Liquid Xenon Ionization and Scintillation Chamber}


 \author[cu]{K. Ni\corauthref{cor}}\ead{nikx@astro.columbia.edu}, \author[cu]{E. Aprile}, \author[cu]{D. Day\thanksref{reu}}, \author[cu]{ K.L. Giboni}, \author[coimbra]{J.A.M. Lopes}, \author[cu]{P. Majewski}, \author[cu]{ M. Yamashita}

\address[cu]{Physics Department and Columbia Astrophysics Laboratory, Columbia University, New York, NY 10027}
\address[coimbra]{Departamento de Fisica, Universidade de Coimbra, P-3004-516 Coimbra, Portugal}
\corauth[cor]{Corresponding author. Tel.: 1-914-591-2825; fax: 1-914-591-4906.} 
\thanks[reu]{Permanent address: University of Florida, Gainesville, FL 32611, Columbia Nevis Lab REU program participant (summer 2004).}

\begin{abstract}
Scintillation light produced in liquid xenon (LXe) by alpha particles, electrons and gamma-rays was detected with a large area avalanche photodiode (LAAPD) immersed in the liquid. The alpha scintillation yield was measured as a function of applied electric field. We estimate the quantum efficiency of the LAAPD to be 45\%. The best energy resolution from the light measurement at zero electric field is 7.5\%($\sigma$) for 976 keV internal conversion electrons from \(^{207}\)Bi and 2.6\%($\sigma$) for 5.5 MeV alpha particles from \(^{241}\)Am. The detector used for these measurements was also operated as a gridded ionization chamber to measure the charge yield. We confirm that using a LAAPD in LXe does not introduce impurities which inhibit the drifting of free electrons. 

\end{abstract}

\end{frontmatter}

\section{Introduction}
\label{}
In recent years, large area avalanche photodiodes (LAAPD) have been successfully applied as photodetectors for a variety of scintillators \cite{Lopes:01,Kapusta:02,Moszynski:03}, including liquid xenon (LXe) \cite{Solovov:02}. LXe scintillation light is in the VUV region, centered at 178 nm \cite{Jortner:65}, which makes it difficult to detect efficiently. Within the context of the R\&D for the XENON dark matter project \cite{XENON}, we are testing different photodetectors directly coupled to LXe, including UV sensitive photomultipliers (Hamamatsu R9288 and R8520), microchannel plate PMTs (Burle 85006), silicon photomultiplier (SiPM) \cite{sipm} and LAAPDs from Advanced Photonix, Inc.(API). In this paper we report our first results on the detection of LXe scintillation with a 16 mm diameter LAAPD, mounted on a custom designed ceramic support \cite{Bolozdynya:pri}. The interest in LAAPDs for LXe is in their high quantum efficiency (QE) at 178 nm, as originally reported by \cite{Solovov:02}. Their negligible radioactive contamination is also attractive for low background experiments based on LXe, such as direct dark matter searches (XENON \cite{XENON} and ZEPLIN \cite{zeplin}) and neutrinoless double beta decay searches (EXO \cite{EXO}).

On the other hand, for a practical application of LAAPDs, especially for LXe based dark matter searches, several issues remain to be addressed. The primary concern is the capability of a light readout based on LAAPDs to achieve the low energy threshold of a few tens of keV required for high sensitivity. Other issues include long term stability of operation, impact on LXe purity, as well as readout complexity and cost. The experiments presented in this paper aimed at confirming the high QE for LXe scintillation reported in the literature, and at verifying the compatibility of LAAPDs immersed in the liquid with the requirement to detect tiny ionization signals, at the single electron level.

\section{Experimental set-up}
\label{}
\begin{figure}
\centering
\includegraphics[width=0.9\textwidth]{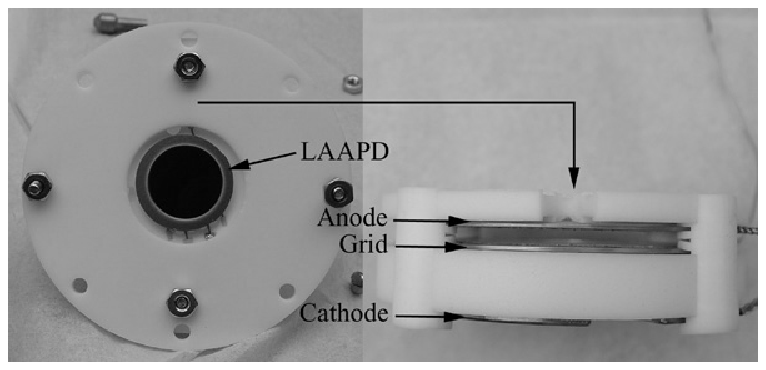}
\caption{Photograph of the assembled gridded ionization chamber electrodes, with the LAAPD mounted 6 mm above the anode grid. A PTFE plate is used to support the LAAPD above the anode grid. PTFE rings are used to maintain the electrodes spacing. }
\label{APD_s}
\end{figure}

The 16 mm diameter, windowless LAAPD tested in these experiments was mounted inside a gridded ionization chamber, filled with high purity LXe. A photograph of the assembled electrodes and LAAPD is shown in Fig.~\ref{APD_s}. The cathode is a
6 cm diameter stainless steel plate with either a gamma ray source
(\(^{207}\)Bi) or an alpha source (\(^{241}\)Am) deposited on its center. The shielding grid and the anode grid are made with stretched wires on stainless steel frames, with a thickness of about 0.8 mm. The wire spacing is 2 mm, and the wire diameter is 60 \(\mu\)m. The separation between cathode and grid is 1 cm, which defines the maximum drift gap for ionization electrons. The grid to anode separation is 3 mm. The electrodes spacing was maintained with rings made of PTFE, for its high reflectivity in the UV region \cite{Yamashita:04}. The LAAPD, originally glued by API on a thin ceramic substrate, was mounted on a PTFE disk, facing the anode grid at a distance of 6 mm above. We note that the LAAPD had been exposed to air for several months prior to its use in LXe. Hermetic feedthroughs are used for cathode and grid HV lines and for anode signal readout via a charge sensitive amplifier. Additional feedthroughs are used to bias the LAAPD and to connect its output to a separate charge sensitive amplifier.

\begin{figure}
\centering
\includegraphics[width=0.9\textwidth]{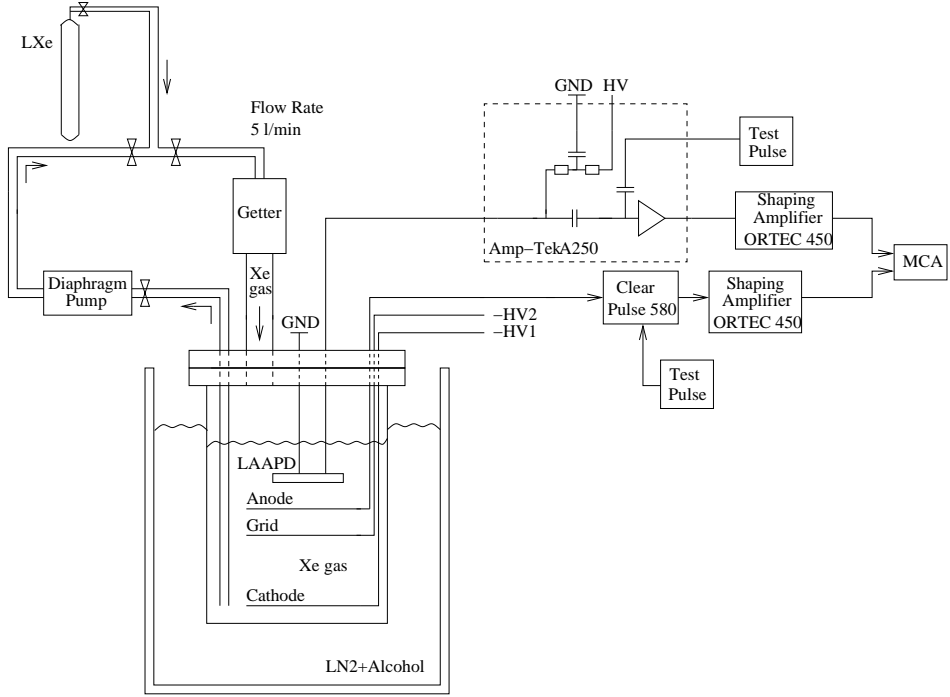}
\caption{DAQ schematics for the LAAPD setup for scintillation and ionization detection in liquid xenon. The top left part shows a simple schematics for the xenon gas recirculation and purification system.}
\label{DAQ}
\end{figure}

A schematic drawing of the detector system and electronics readout is shown in Fig.~\ref{DAQ}. The ionization
electrons, which are created from gamma rays or alpha particles in the drift
region, are drifted upward by the applied electric field, and are collected by a charge
sensitive pre-amplifier (ClearPulse Model 580 \cite{ClearPulse}) connected to the anode. The scintillation light hits the LAAPD and
produces photoelectrons, which are amplified by the avalanche process. The avalanche electrons are collected by an AmpTek 250
pre-amplifier \cite{AmpTek}. The charge and light signals, either from ClearPulse 580 or
AmpTek A250, are fed into a low-noise shaping amplifier (ORTEC 450 \cite{ORTEC}). The amplified signals are further fed into a PC-based multi-channel analyser (MCA) for spectroscopy analysis. Known test
pulses are used to calibrate the DAQ system, for both light and charge
signals. The capacitances in the pre-amplifiers were calibrated with a silicon detector. We used an open bath cooling apparatus with a liquid nitrogen and alcohol mixture to condense the xenon gas. The vessel enclosing the assembled detector was filled with high purity LXe, covering completely the LAAPD. As shown in Fig. \ref{DAQ}, we used a high temperature getter to purify Xe gas continuously via a diaphragm pump. The system, developed for XENON prototypes, is described in \cite{dm2004}.

\section{Experimental measurements}
\label{}
\subsection{LAAPD gain}
\begin{figure}
\centering
\includegraphics[width=0.9\textwidth]{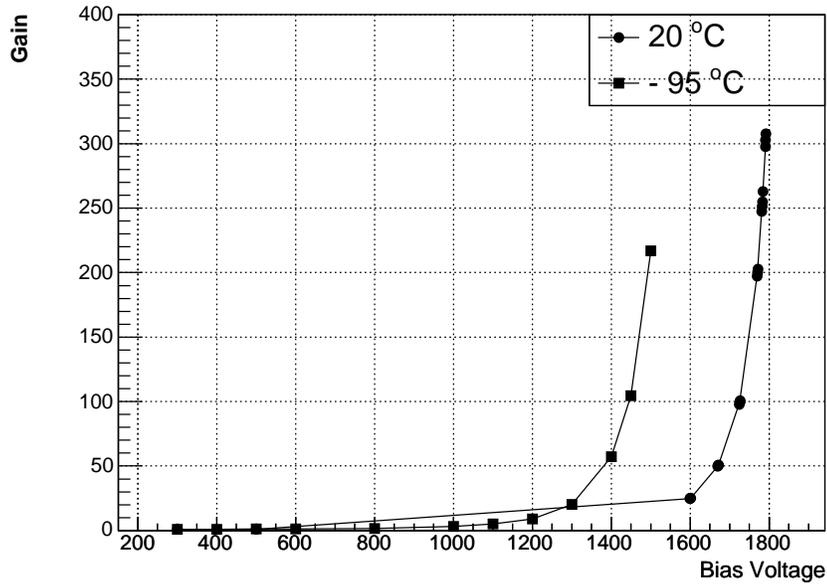}
\caption{LAAPD gain as a function of bias voltage measured at liquid xenon (-95\(\rm ^oC\))
and room temperature (20\(\rm ^oC\)). The data at room temperature is obtained from \cite{Bolozdynya:pri}. The gain with bias voltage of 1400 V is 57 at -95\(\rm ^oC\).}
\label{gain}
\end{figure}

The gain of the LAAPD was measured in a different set-up configuration, which optimizes the light collection efficiency by placing the LAAPD very close (about 3.5 mm) to the source plate, at -95\(\rm ^oC\) (Fig.~\ref{gain}). The direct scintillation from a 5.5 MeV alpha source (\(^{241}\)Am) was measured as a function of applied voltage on the LAAPD up to about 1500 V. Unitary gain was determined from the average amplitude in the 300 to 500 V range \cite{Fernandes:04a}. The typical reduction in applied voltage, for a constant gain, when reducing temperature \cite{Solovov:00} was observed, corresponding to an average voltage variation of about 2.4 V/\(\rm ^oC\) at a gain of 100.

\subsection{Scintillation spectroscopy}
\begin{figure}
\centering
\includegraphics[width=0.9\textwidth]{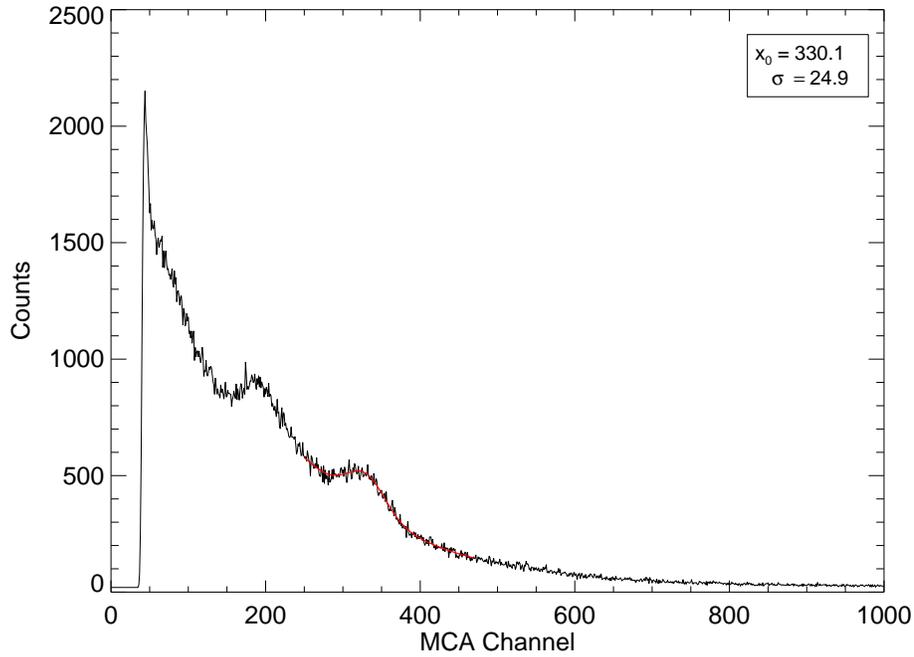}
\caption{Scintillation light spectrum from \(^{207}\)Bi at zero electric field. LAAPD gain is about 146 and the temperature is -95\(\rm ^oC\). The peak of the 976 keV electron line is fitted at about channel 330, giving a resolution of 7.5\% (\(\sigma\)). The 570 keV gamma line can also be seen clearly at about channel 190.}
\label{bi-spe}
\end{figure}

The scintillation light spectrum of the \(^{207}\)Bi radiation in LXe was measured, as
shown in Fig.~\ref{bi-spe}. Due to the small liquid xenon sensitive volume, the energy from most of the 1064 keV gamma rays of \(^{207}\)Bi is not fully deposited. The second peak on the spectrum is mostly contributed by the 976 keV internal conversion electrons. As the gamma rays interact at any point in the liquid xenon active volume, light collection in the LAAPD varies accordingly to the interaction position. Using PTFE reflectors, the variation in light collection can be reduced to less than 1\%, without compromising the energy resolution. The energy resolution for the 976 keV electrons of \(^{207}\)Bi is 7.5\% (\(\sigma\)), which is comparable to the energy resolution obtained earlier by using a PMT in the same chamber with similar geometry \cite{Aprile:04}. The spectrum was accumulated at zero electric field to maximize light output from LXe.

\begin{figure}
\centering
\includegraphics[width=0.9\textwidth]{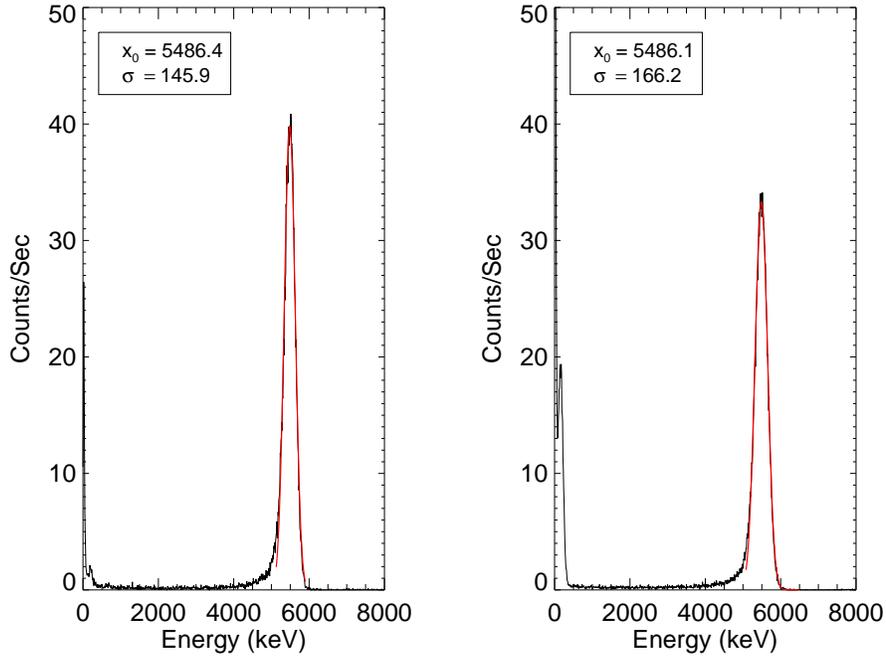}
\caption{\(^{241}\)Am scintillation light energy spectrum from 5.5 MeV alpha particles detected by the LAAPD in LXe. The LXe temperature is -95\(\rm ^oC\) and LAAPD gain is 57. The alpha peak is fitted with a gaussian function. The energy resolution is 2.6\% for the case with the PTFE wall (left) and 3.0\% for the case without the PTFE wall (right). The peaks are normalized to the energy of the alpha particles.}
\label{alpha-spe}
\end{figure}

To better evaluate the LAAPD performance in liquid xenon, larger amounts of scintillation light are necessary. The 5.5 MeV alpha particles from \(^{241}\)Am provide typically one order of magnitude larger scintillation pulse and their interaction is very localized in liquid xenon, contributing to a clear scintillation light spectrum. To get the best possible energy resolution, the gain of LAAPD and shaping time on the amplifier were optimized and finally a gain of 57 was selected for this measurement as shown in Fig.~\ref{alpha-spe}. A very good energy resolution of 2.6\% ($\sigma$) (with PTFE walls) and 3.0\% ($\sigma$) (without PTFE walls) was obtained. To compare the performance of LAAPD with other photon detection devices, the alpha spectrum was also measured using a 2-inch diameter Hamamatsu R9288 PMT with less than 20\% QE at 178 nm wavelength. A value of 2.5\% energy resolution was obtained.

The energy resolution as a function of LAAPD gain can be written as follows,

\begin{center}
\begin{equation}
\sigma/E = \sqrt{\big(\frac{N_e}{N_0 M}\big)^2 + \frac{F-1}{N_0} + \delta^2}
\end{equation}
\end{center}

\begin{figure}
\centering
\includegraphics[width=0.9\textwidth]{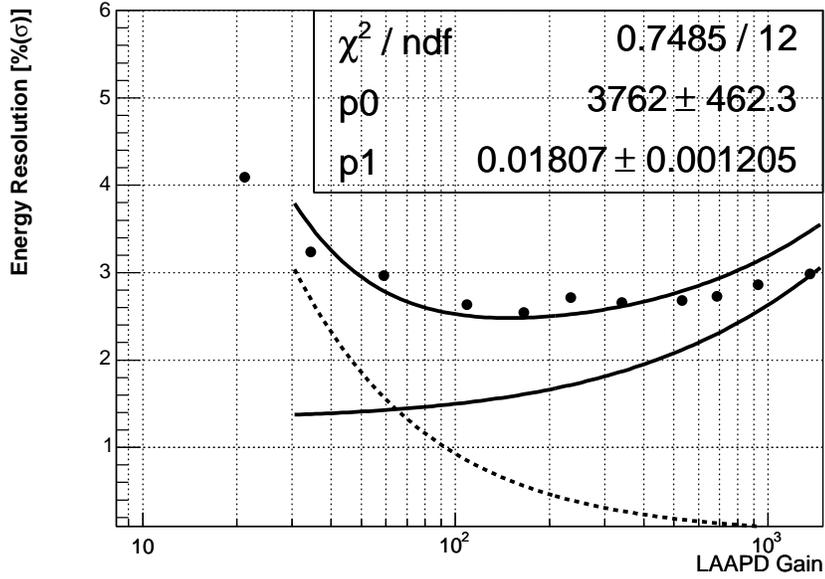}
\caption{Energy resolution as a function of LAAPD gain (thick line with experimental data points). The fitting parameter p0 is the noise equivalent charge \(N_e\), and p1 is \(\delta\) in eq.1. The noise contributions from electronic noise (dashed line, first term in eq.1) and excess noise factor term (thick line without data points, second term in eq.1), from the fitting parameters, as functions of LAAPD gain are also plotted. }
\label{res_vs_g}
\end{figure}

The first term in this equation is contributed from electronic noise. \(N_e\) is
the noise equivalent charge, which can be measured from the distribution of a
known test pulse. \(N_0\) is the number of primary electron-hole pair produced from the
scintillation light in the photodiode surface. \(M\) is the LAAPD gain. The
second term is from the fluctuations of the gain and is inherent to the electron
avalanche process of the LAAPD, where \(F\) is the
excess noise factor. The experimental value for $F$ is approximately written as $F=1.7003 + 0.0021M$ \cite{Fernandes:04b} for $M>30$. The third term \(\delta\) is contributed from the scintillation
process in the liquid xenon, including statistical fluctuations of scintillation photon production, and photoelectrons created in the LAAPD. Using the gain measurement with the LAAPD at 3.5 mm from the source plate, we fit the results with the above equation with  $N_e$ and $\delta$ as free parameters (Fig.~\ref{res_vs_g}), while fixing $F$ from the experimental value \cite{Fernandes:04b}. The noise equivalent charge $N_e$ from the fit agrees well with the measured value with a calibrated test pulse. From the fitted values, we infer that the statistical fluctuations contribute 1.8\% to the energy resolution, with a contribution of about 1.6\% ($1/\sqrt{N_0}$, $N_0 \approx 4000$) from fluctuations in the number of photoelectrons.

\section{Quantum efficiency}
\label{qe}
The quantum efficiency (\(\eta\)) of a LAAPD can be calculated by the following
equation,

\begin{center}
\begin{equation}
\eta = \frac{N_0}{N_p} = \frac{N_d/M}{\alpha N_{tot}}
\end{equation}
\end{center}

Here \(N_0\) is the number of photoelectrons from the LAAPD, \(N_p\) is the number of photons reaching
the LAAPD surface, $M$ is the LAAPD gain,  $N_d$ is the number of electron charges detected by the
pre-amplifier and $N_{tot}$ is the total number of scintillation
photons produced by an event. \(N_{tot}\) is approximately equal to \(E/W_{ph}\), where \(E\) is the energy of the
event, and \(W_{ph}\) is the average energy required to produce a scintillation photon in liquid xenon. The
$W_{ph}$ values are 21.6 eV \cite{Doke:02} and 19.6 eV \cite{Doke:99} for gamma and alpha events respectively. \(\alpha\) is the light collection efficiency,
which is defined as the percentage of the total LXe light yield reaching the LAAPD surface. We have estimated the light collection efficiency by
using a light tracing simulation program in GEANT4 \cite{GEANT4} with the assumptions listed in Table 1.

The estimated light collection efficiency from the simulation is 7.0 $\pm$ 0.7\%
for the structure with 1 cm thick PTFE wall between the cathode and grid. The error
indicates the different reflectivity values used in the simulations. In the case of no PTFE walls, the light collection efficiency was calculated by simply using the solid angle for alpha particles and ignoring the reflectivity of stainless steel. As result, 3.3\%  light collection efficiency was obtained. Considering a 20\% stainless steel reflectivity, 4.1\% light collection efficiency was obtained. In order to estimate the number of photons incident on the LAAPD surface, an average value for light collection efficiency was used, as presented in Table 2.

From the above considerations, we estimated the QE of the LAAPD
for different measurements, which are shown in Table 2. The main uncertainty is from the estimation of the light collection efficiency. The different values obtained from electron and alpha events may be due to the uncertainty in $W_{ph}$ and the LAAPD gain values. In conclusion, we use the average of the QE values from the electron and alpha measurements, which is $45\pm5\%$.

\section{Field dependence of light yield}
\label{}
\begin{figure}
\centering
\includegraphics[width=0.9\textwidth]{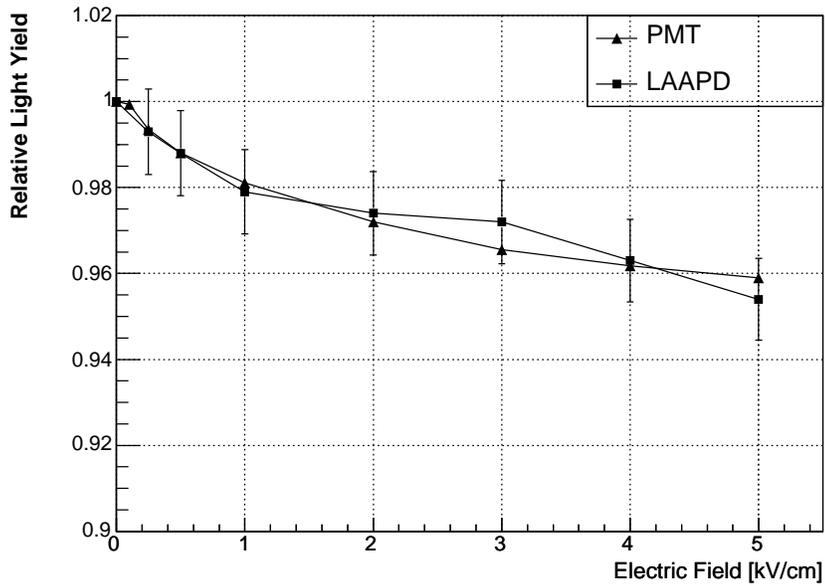}
\caption{Field dependence of light yield for \(^{241}\)Am 5.5 MeV alpha particles. The measurements were carried out with either a PMT or a LAAPD, immersed in liquid xenon at around -95\(\rm ^oC\). The error bars on the LAAPD data are from the LAAPD gain variation due to temperature fluctuations.}
\label{field_dep}
\end{figure}

The liquid xenon scintillation light yield depends on the strength of the applied electric field \cite{Doke:02}. The dependence of the scintillation yield of liquid xenon for alpha particles has been measured by the Columbia group several years ago, using an external PMT coupled to a LXe volume via a CaF\(_2\) window \cite{Aprile:90}. In the current setup, we were able to measure this field dependence with a LAAPD. For comparison, we also measured this field dependence with a Hamamatsu R9288 PMT immersed in liquid xenon, in the same chamber. Fig.~\ref{field_dep} shows the combined results.

The LAAPD gain varies considerably
 with temperature, and our simple alcohol-LN\(_2\) cooling bath does not keep the liquid temperature stable enough to avoid temperature dependent gain variations. The gain of the PMT is not much affected by such small temperature fluctuations. The data from the LAAPD and the PMT are in good agreement, but the result with the LAAPD has more fluctuations due to its acute temperature dependence, which is obvious from the curve.

\section{Impact of LAAPD on LXe purity}
\label{}
\begin{figure}
\centering
\includegraphics[width=0.9\textwidth]{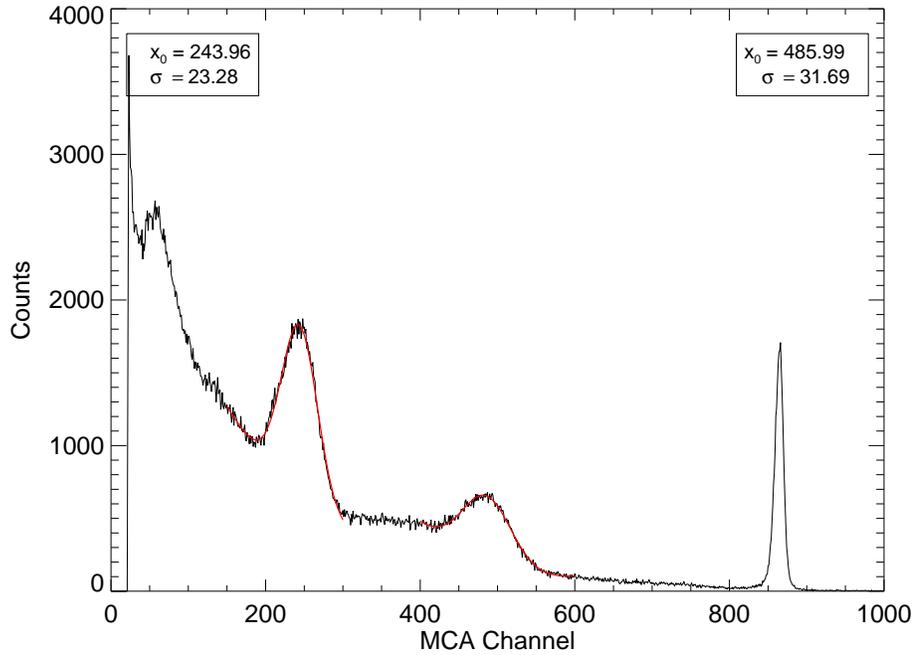}
\caption{Ionization spectrum from \(^{207}\)Bi radiation in liquid xenon at 1 kV/cm drift field with an LAAPD immersed in the liquid. The peaks are from 570 keV gamma rays (channel 244) and 976 keV internal conversion electrons (channel 486). The sharp peak on the right of the charge spectrum is from a known test pulse. Based on the peak values, we estimated the charge collection at about 75\%, by assuming the average energy needed to produce an electron ion pair in liquid xenon is 15.6 eV \cite{Takahashi:75}. The energy resolution for the 570 keV gamma rays and 976 keV electron peaks are 9.5\% and 6.5\% separately.}
\label{bi_qspe}
\end{figure}

One challenge involving photon detection devices immersed in liquid xenon is their compatibility with the high purity required for electron drift, if a combined charge and light readout is implemented. Our experience with LXe detectors shows that the light yield of LXe is not very sensitive to the purity level, unlike the charge yield. Many efforts have been made in the past to detect both ionization and scintillation, using PMTs, in liquid xenon \cite{Aprile:03}. Currently
we have developed a xenon recirculation and purification system
\cite{dm2004}, which continuously removes impurities from the liquid
xenon during experiment. Depending on the size of
the detector, we can achieve a sufficient purity level for drifting ionization
electrons within two to tens of hours. During the first experiment with the LAAPD immersed in the LXe, we used this recirculation system to purify the xenon, continuously. We measured
the 976 keV peak position of the \(^{207}\)Bi to monitor the charge
collection. Within a few hours of recirculation and
purification, we achieved a high charge collection of 75\%. The ionization spectrum of \(^{207}\)Bi at 1 kV/cm drift field shown in Fig.~\ref{bi_qspe} is comparable to that measured in a liquid xenon ionization chamber with an external PMT \cite{Aprile:91}. From the second
experiment, we observed the same level of charge collection even without using the recirculation system, which indicates that the LAAPD is clean and does not bring any impurities into the liquid xenon.

\section{Conclusion}
\label{}
In this paper, we have demonstrated the operation of a LAAPD in liquid xenon to detect scintillation light from gamma rays and alpha particles. We have achieved the best energy resolution, for alpha particles and fast electrons, from liquid xenon scintillation light detected by a LAAPD. The inferred value of the QE is 45\%, which is lower than that previously reported in \cite{Solovov:02}. A recent article also reports a higher QE value from a measurement in gas xenon \cite{Chandrasekharan:04}. We have repeated the measurements with the LAAPD sample and confirm that the results are reproducible. A possible explanation for the lower QE is the intrinsic quality of the LAAPD. This explanation appears consistent with the subsequent QE measurement of the same LAAPD at room temperature, using an indepedent setup. The QE was inferred from the ratio of the current output from the LAAPD and a calibrated PIN diode, by irradiating UV light from a xenon lamp to the devices \cite{Shagin:pri}. The measured value ($39\pm 3\%$) is very close to what we measured in liquid xenon. We believe that the QE of this device should not change significantly with temperature. We intend to test a new LAAPD of the same type as used in \cite{Chandrasekharan:04} to investigate if the QE we measured was specific to the sensor used.

The gain of the LAAPD is much lower than that of PMTs. Using an external amplification system limits the energy detection threshold, which is very crucial for a sensitive dark matter detector. The scintillation light from a WIMP recoil of a few tens of keV energy may produce not enough light to be detected by the LAAPD. On the other hand, properties of LAAPDs, such as their compact size, very high QE, and compatibility with low radioactive background and high LXe purity requirements, make them attractive for applications of LXe detectors in particle physics, medical imaging and astrophysics.

\section{Acknowledgments}
\label{}
This work was supported by a grant from the National Science Foundation to the Columbia Astrophysics Laboratory (Grant No. PHY-02-01740). One of the authors (P. Majewski) acknowledges the support by the North Atlantic Treaty Organization under a grant awarded in 2003.

\begin{center}
\begin{table}{\bfseries Table 1: Physical parameters for the light collection simulation}\\
\begin{tabular}{ccc}
\hline
Item&Value&Ref.\\
\hline
{PTFE reflectivity}&{90-95\%}&{\cite{Yamashita:04}}\\
{Light absorption length}&{100 cm}&{\cite{Baldini:04}}\\
{Rayleigh scattering length}&{30 cm}&{\cite{Seidel:02}}\\
\hline
\end{tabular}
\end{table}
\end{center}

\begin{center}
\begin{table}{\bfseries Table 2: LAAPD quantum efficiency measured from different sources}\\
\begin{tabular}{lccccl}
\hline
Measurement&PTFE&Light Col. Eff.(\%)&Bias V&APD Gain&QE(\%)\\
\hline
976 keV e$^-$&with&{7.0 $\pm$ 0.7}&1460&{146}&{50 $\pm$ 5}\\
{5.5 MeV \(\alpha\)}&with&{7.0 $\pm$ 0.7}&1400&{57}&{39 $\pm$ 4}\\
{5.5 MeV \(\alpha\)}&without&{3.7 $\pm$ 0.4}&1400&{57}&{39 $\pm$ 4}\\
\hline
\end{tabular}
\end{table}
\end{center}

\end{document}